\documentclass[twocolumns]{aa}
\usepackage[dvips]{graphicx}
\usepackage{txfonts}
\newcommand{\be}{\begin{equation}}
\newcommand{\ee}{\end{equation}}
\newcommand{\bea}{\begin{eqnarray}}
\newcommand{\eea}{\end{eqnarray}}
\begin{document}

\title{Secular dynamics of planetesimals in tight binary systems: Application 
to $\gamma$-Cephei}

\author{C.A. Giuppone$^{1}$ \and A.M. Leiva$^{1,2}$ \and J. Correa-Otto$^{1,2}$ 
\and C. Beaug\'e$^{1,2}$}

\institute{Observatorio Astron\'omico, Universidad Nacional de C\'ordoba, 
Laprida 854, (X5000BGR) C\'ordoba, Argentina \\
\email{cristian@oac.uncor.edu} 
\and Instituto de Astronom\'ia Te\'orica y Experimental, Laprida 854, (X5000BGR) C\'ordoba, Argentina}

\titlerunning{Secular dynamics of planetesimals in tight binary systems}
\authorrunning{Giuppone et al.}

\abstract{The secular dynamics of small planetesimals in tight binary systems play a fundamental role in establishing the possibility of accretional collisions in such extreme cases. The most important secular parameters are the forced eccentricity and secular frequency, which depend on the initial conditions of the particles, as well as on the mass and orbital parameters of the secondary star.}
{We construct a second-order theory (with respect to the masses) for the planar secular motion of small planetasimals and deduce new expressions for the forced eccentricity and secular frequency. We also reanalyze the radial velocity data available for $\gamma$-Cephei and present a series of orbital solutions leading to residuals compatible with the best fits. Finally, we discuss how different orbital configurations for $\gamma$-Cephei may affect the dynamics of small bodies in circunmstellar motion.}
{The secular theory is constructed using a Lie series perturbation scheme restricted to second order in the small parameter. The orbital fits were analyzed is done with a minimization code that employs a genetic algorithm for a preliminary solution plus a simulated annealing for the fine tuning.}
{For $\gamma$-Cephei, we find that the classical first-order expressions for the secular frequency and forced eccentricity lead to large inaccuracies {$\sim$ 50\%} for semimajor axes larger than {one tenth} the orbital separation between the stellar components. Low eccentricities and/or masses reduce the importance of the second-order terms. The dynamics of small planetesimals only show a weak dependence with the orbital fits of the stellar components, and the same result is found including the effects of a nonlinear gas drag. Thus, the possibility of planetary formation in this binary system largely appears insensitive to the orbital fits adopted for the stellar components, and any future alterations in the system parameters (due to new observations) should not change this picture. Finally, we show that planetesimals migrating because of gas drag may be trapped in mean-motion resonances with the binary, even though the migration is divergent.}{}


\maketitle

\date

\section{Introduction}

Although it is believed that approximately half of all stars belong to multiple stellar systems (e.g. Duquennoy and Mayor 1991), $\sim 90 \%$ of exoplanets are associated with single stars (Zsom et al. 2010). It is not yet clear whether this discrepancy is solely due to observational bias, or if the process of planetary formation may be seriously impaired even in very wide binary systems. Curiously, however, a few exoplanets have also been detected in very tight binary stellar systems, where the gravitational perturbations of the secondary component are so large that accretional collisions among small planetesimals are extremely difficult. 

Perhaps the most extreme case is $\gamma$-Cephei, a binary stellar system whose most recent orbital determination (Neuh\"auser et al. 2007) shows a secondary component of mass $m_B \sim 0.4 M_{\odot}$ orbiting a principal star $m_A \sim 1.4 M_{\odot}$ in an ellipse with semimajor axis $a_B \sim 20.2$ AU and eccentricity $e_B \sim 0.41$. Although both stars have a minimum mutual distance of only $\sim 12$ AU, a giant planet has been detected at $\sim 2$ AU from $m_A$ (Hatzes et al. 2003). So far, all attempts to understand the accretional history of this extrasolar planet have been unsuccessful, and it is difficult to visualize an scenario under which such a massive Jovian planet could form through accretional collisions from a primordial planetesimal swarm. 

The $\gamma$-Cephei system then constitutes a paradigm. It may be argued that if we are able to comprehend planetary formation in such an extreme environment, we would have taken large steps towards a global understanding of planetary formation in any other system. It is then no surprise that this planetary system has caught the attention of several researchers over the past decade, and many dynamical and collisional studies may be found in the literature (e.g. Th\'ebault et al. 2004, 2006, Haghighipour 2006, Verrier and Evans 2006, Tsukamoto and Makino 2007, Paardekooper et al. 2008, Xie and Zhou 2008, Kley and Nelson 2008, Paardekooper and Leinhardt 2010). 

In Beaug\'e et al. (2010) we showed that the dynamics of the gas disk plays an important role in determining the evolution of small planetesimals. We found that, although an apsidal precession of the gas elements may play a disruptive role, especially in the inner parts of the disk, the combined effects of a non-zero precession rate plus a high forced eccentricity in the disk may in fact lower the relative velocity of solid bodies in the outer regions. We thus envisioned a possible scenario in which the growth from kilometer-size planetesimals to $\sim 100$ km planetary embryos could initially take place near the truncation radius of the gas disk. As the embryos spiral down towards a lower semimajor axis, subsequent collisions could then lead to larger embryos, finally forming a giant planet core hopefully near the present location of the currently detected Jovian planet. 

To test this idea, we performed preliminary N-body simulations of the dynamical and collisional evolution of planetesimal swarms in the outer regions of the gas disk. However, it was soon realized that the dynamics of the solid bodies in this region is extremely complex. First, we found that mean-motion resonances with the secondary star have significant effects, even though they are of very high order (e.g. 12/1, 11/1, 10/1). Second, the proximity to the secondary star also affects the secular dynamics, and the classical analytical models (e.g. Heppenheimer 1978) widely used in these problems fail to reproduce the correct orbital variations. The imprecision does not lie in the reduced expression adopted for disturbing potential, but in the averaging process used to eliminate short-period terms. Similar to the irregular satellites around the outer planets of our solar system (e.g. \'Cuk and Burns 2004, Beaug\'e et al. 2006), higher order secular effects from the interaction of short-period terms (including evection) must be considered to represent the dynamics of planetesimals in close binary systems. However, contrary to the satellite problem, here the perturber lies in a high-eccentricity orbit, and classical high-order models cannot be applied directly (Correa Otto et al. 2010).

The purpose of this work is to present a general description of the secular dynamics of small planetesimals in circumstellar motion around the more massive star of a generic binary stellar system. {We assume that all bodies share the same orbital plane. Although it is known that even moderate mutual inclinations can have significant effects in the accretional evolution of a planetesimal swarm (e.g. Marzari et al 2009a, Xie and Zhou 2009, Xie et al 2010, Fragner et al 2011), here we concentrate on the planar case and leave the extension to 3D for a future work.} Finally, although our model will be generic, we apply the results to the particular case of $\gamma$-Cephei where we analyze how the uncertainties in the orbital fits of the secondary stellar companion may affect the evolution of a planetesimal swarm.

The paper is organized as follows. Section 2 presents our second-order perturbation model and analytical approximations for both the forced eccentricity and secular frequency. Since we focus our attention on $\gamma$-Cephei, Section 3 discusses the orbital parameters determined for the two stellar components and their precision. The secular dynamics of individual planetesimals, under the additional effects of a nonlinear gas drag, is analyzed in Section 4. We also present an example of resonance trapping obtained with divergent migration. Finally, discussions close the paper in Section 5.

\section{Secular dynamics}

Let us assume a small planetesimal of mass $m$ in circumstellar motion around a star of mass $m_A$, which is in turn part of a binary system with a smaller component of mass $m_B$. Let $a_B$ be the $m_A$-centric semimajor axis of $m_B$ and $e_B$ its orbital eccentricity. We further assume that all motion occurs in a plane. 

Neglecting the gravitational effects of $m$ on both stellar bodies, the orbit of $m_B$ will be a fixed ellipse, while the motion of the small planetesimal will be perturbed by the gravitational effects stemming from the secondary component. Thus, in our dynamical system, $m_B$ will play the role of the perturber, while $m$ will be the perturbed mass.

\subsection{The first-order secular model}

Outside any significant mean-motion resonance, the orbital evolution of $m$ will be dominated by the secular perturbations, and the short-period terms (associated to the mean longitudes) can be eliminated by a perturbation technique known as {\it averaging}. The expression for the secular disturbing function $R$ usually employed for these studies was originally developed by Heppenheimer (1978) which, except for constant terms, is given by
\be
\label{eq1}
R = \frac{3}{8}\frac{{\cal G} m_B}{(1-e_B^2)^{3/2}} \frac{a^2}{a_B^3}
      \biggl[ e^2 - \frac{5}{2} \frac{a e e_B}{a_B(1-e_B^2)} 
       \cos{(\varpi-\varpi_B)} \biggr]
\ee
(see Terquem and Papaloizou 2002), where ${\cal G}$ is the gravitational constant, $a$ is the $m_A$-centric semimajor axis of the planetesimal, $e$ its eccentricity, and $\varpi$ its longitude of pericenter. The angle $\varpi_B$ denotes the longitude of pericenter of the orbit of $m_B$, assumed constant. 

Expression (\ref{eq1}) is constructed from Kaula's (1962) expansion of the disturbing potential, truncated to second-order expansion in the eccentricity of the perturbed body, and performing a first-order ``scissors'' averaging (with respect to the masses) in the mean longitudes. We refer to the resulting expressions as a {\it first-order} model for the secular dynamics.

Since $R$ does not depend explicitly on the mean longitude $\lambda$ of the planetesimal, its semimajor axis is constant and equal to the proper value $a^*$. Consequently, the secular system is reduced to a single degree of freedom, and the differential equations governing the regular variables $k=e \cos{(\varpi-\varpi_B)}$ and $h=e \sin{(\varpi-\varpi_B)}$ can be written as
\be
\label{eq2}
\frac{dk}{dt} = -g h \hspace*{0.5cm} ; \hspace*{0.5cm}
\frac{dh}{dt} =  g (k - e_f),
\ee
where
\bea
\label{eq3}
g &=& \frac{3}{4} \frac{m_B}{m_A} \biggl( \frac{a}{a_B} \biggr)^3 n(1-e_B^2)^{-3/2} \\
\label{eq4}
e_f &=& \frac{5}{4} \frac{a}{a_B} e_B (1-e_B^2)^{-1} .
\eea

Given arbitrary initial conditions $(k_0,h_0)$, these equations admit periodic solutions of the form
\bea
\label{eq5}
k(t) & = & e_p \cos{(g t + \phi_0)} + e_f  \\
h(t) & = & e_p \sin{(g t + \phi_0)}        \nonumber,
\eea
where $g$ is the secular frequency, $e_p^{2}=(k_0-e_f)^{2}+h_0^{2}$ is usually known as the proper (or free) eccentricity, and the phase angle is given by the expression $\tan \phi_0=h_0/(k_0-e_f)$. The constant term $e_f$ is known as the forced eccentricity and is only present in systems with an eccentric perturber. Adopting fixed values for $a_B$ and $e_B$, equation (\ref{eq4}) implies that $e_f$ is a linear function of the proper semimajor axis ($e_f \sim a^*$) while the secular frequency scales as $g \sim {a^*}^{3/2}$.

\subsection{Numerical simulations}

Our first task is to assess the accuracy of the secular solutions (\ref{eq5}) corresponding to the first-order model (\ref{eq1}). Two quantities we particularly wish to test are $e_f$ and $g$. The forced eccentricity is crucial in determining the equilibrium eccentricity of planetesimals under the effects of gas drag from the protoplanetary nebula. Although any secular oscillatory motion is expected to be damped in a gas-rich scenario, the magnitude of $g$ is important for establishing the validity of the averaging process of the disturbing function. 

For our computations, we assume a generic binary system with mass ratio between the components of $m_B/m_A = 0.4$ and eccentricity $e_B=0.36$. This value is similar to the best-fit solution found by Hatzes et al. (2003) for $\gamma$-Cephei. 

\begin{figure}
\centerline{\includegraphics*[width=0.95\columnwidth]{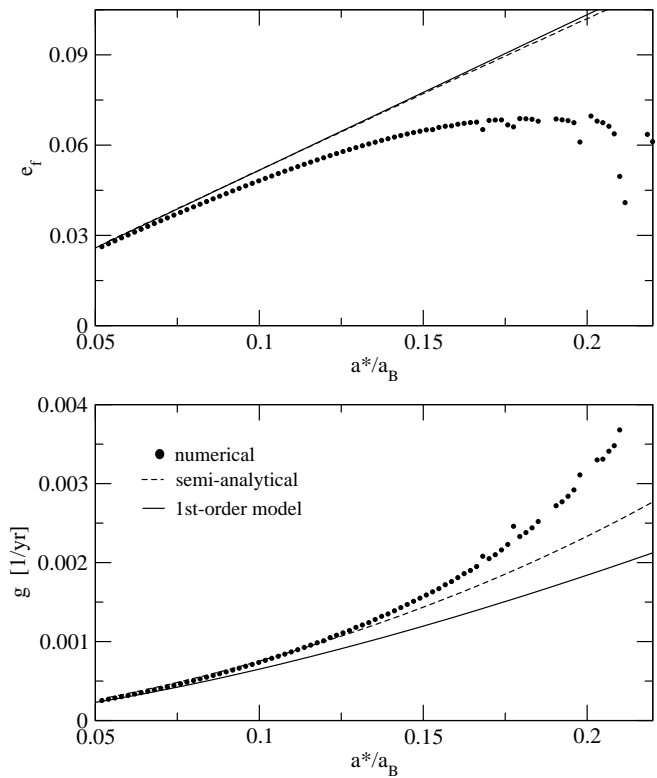}}
\caption{Forced eccentricity (top) and secular frequency (bottom), as function of the proper semimajor axis, calculated by three different methods: filtered exact numerical simulations (filled black circles), semi-analytical first-order averaging of the exact disturbing function (dashed lines), and the classical analytical first-order secular model {using Eqs. \ref{eq3} and \ref{eq4}} (continuous lines).}
\label{fig1}
\end{figure}

Figure \ref{fig1} shows three different calculations of the forced eccentricity (top frame) and the secular frequency (bottom frame). 
The value of $e_f$ appears to grow linearly with the proper semimajor axis, reaching values of $\sim 0.1$ for $a^* \sim 0.2 a_B$. The numerical calculations were obtained from a long-term integration of 
the exact equations of motion, after an online application of a low-pass FIR {(finite impulse response)} filter (e.g. Carpino et al. 1987). 

A digital filter is a numerical tool that eliminates certain frequencies from an input signal. For example, given a certain time series (e.g. eccentricity as function of time) and a pass frequency $\nu_{\rm pass}$,  applying a low-pass filter signal will yield an output that maintains all the periodic variations with frequencies $\nu < \nu_{\rm pass}$ while eliminating the rest. Digital filters are a common tool for constructing synthetic theories of long-term asteroid dynamics (e.g. Knezevi\'ic and Milani 2000) and planetary dynamics (e.g. Michtchenko and Ferraz-Mello 2001), and constitute a useful alternative to analytical perturbation theories when the Hamiltonian function is very complex.

For the present work, the parameters of the digital filter were chosen to eliminate all periodic variations with up to eight times the orbital period of the binary component. In a dynamical system displaying regular motion, the application of the filter is equivalent to a full (i.e. infinite-order) averaging of the Hamiltonian. The region located beyond $a^* \sim 0.2 a_B$ shows dynamical instabilities that complicate the determination of the secular solution. 

A comparison between the numerical and the analytical values shows significant differences. Although both methods yield similar results for low values of the semimajor axis, the exact secular frequencies are systematically underestimated by the analytical model, leading to differences of almost a factor of two for $a^*/a_B \sim 0.24$. This limitation in the classical estimation of $g$ was first noticed by Th\'ebault et al. (2006), who presented an empirical functional correction term to expression (\ref{eq3}). This correction reduced the discrepancy to values of around $5 \%$ in the same range of $a^*$. 

Perhaps more important is that the value of the forced eccentricity also shows significant differences. While the analytical model predicts a monotonic increase in $e_f$ as function of $a^*$, the real value appears to reach a plateau around $a^*/a_b \simeq 0.17$ (corresponding to $e_f \simeq 0.07$) and decrease for larger radial distances. The scatter in the numerical values of both $e_f$ and $g$ in this outer region stems from the action of high-order mean-motion resonances between the massless body and the binary star $m_B$. 

At first hand, it seems natural to believe that the limitations of expressions (\ref{eq3})-(\ref{eq4}) are due to the truncation of the disturbing function to second-order terms in the eccentricities and/or third-order terms in the ratio $a/a_B$. However, this is not the case. In Figure \ref{fig1} we have also plotted the same quantities determined using a semi-analytical model for the disturbing function. This expression is calculated directly as
\be
\label{eq6}
\langle R \rangle = \frac{{\cal G} m_B}{(2\pi)^2} \int_0^{2\pi} \int_0^{2\pi}
\biggl( \frac{1}{|{\bf r} - {\bf r_B}|} - \frac{r}{r_B^2} \cos{\phi} \biggr) 
 d\lambda d\lambda_B,
\ee
where ${\bf r}$ and ${\bf r_B}$ are the position vectors of $m$ and $m_B$, respectively, $r$ and $r_B$ are their absolute values, $\lambda$ and $\lambda_B$ are the mean longitudes, and $\phi$ is the instantaneous angular distance between both bodies. The integrand is the exact expression for the disturbing function with no approximations, and the double integral is performed numerically. From this expression the value of $e_f$ can be estimated from the minimum value attained by $\langle R \rangle$ in the line segment $(\varpi-\varpi_0)=0$, while the secular frequency is given by 
\be
\label{eq7}
g = \frac{\partial \langle R \rangle }{\partial (G-L)}
\ee
at the same point. Here $(G-L) \simeq \sqrt{{\cal G}m_A a} \; (e^2/2)$ is the modified Delaunay canonical momenta conjugate to the longitude of the pericenter. Expression (\ref{eq7}) may also be evaluated numerically for any initial condition. 

This type of semi-analytical model has proved to be a powerful tool for mapping the phase space of complex dynamical systems, especially in the high-eccentricity regime where analytical approximations for the Hamiltonian are not available (e.g. Michtchenko and Malhotra 2004, Michtchenko et al. 2006). Formally, it is equivalent to a first-order averaging (in the masses) of the exact Hamiltonian function. 

Figure \ref{fig1} shows the values of both $e_f$ and $g$ determined with this semi-analytical approach. Although the value of the secular frequency shows a significant improvement over the analytical estimation, there is still a discrepancy with the exact value. This is even more noticeable in the forced eccentricity, where there is practically no difference with the value determined from equation (\ref{eq3}). Consequently, it appears that the limitations of the analytical model are not due primarily to the truncation of the disturbing function.

\subsection{A second-order secular model}

Since the errors in the estimation of both the forced eccentricity $e_f$ and the secular frequency $g$ do not come from the limitations in the adopted disturbing function, their origin must lie in the construction of the averaged solution itself. As mentioned previously, Heppenheimer's (1978) expressions are a first-order model with respect to the perturbing mass. Here, we extend the calculations to the second order. 

One of the most widely used perturbation techniques is the so-called Hori's averaging process (Hori 1966, see also Ferraz-Mello 2007), which employs Lie-type canonical transformations to eliminate the dependence of the Hamiltonian with respect to a given set of variables. The new Hamiltonian function is given by a power series in the small parameter (e.g. perturbing mass). 

Since we adopt a Hamiltonian formulation, we first need to introduce canonical variables. We have chosen the modified Delaunay variables $(L,\Lambda,G-L,\lambda,\lambda_B,\varpi)$, where the canonical momenta are given in terms of the orbital elements, by
\be
\label{eq8}
L = \sqrt{{\cal G} m_A a} \hspace*{0.5cm} ; \hspace*{0.5cm}
G-L = L (\sqrt{1-e^2} - 1)
\ee
and $\Lambda$ is the canonical conjugate of the mean longitude of the perturbing mass (i.e. $\lambda_B$). This third degree of freedom appears when passing to the extended phase space to eliminate the non-autonomous character of the perturbation.

The full Hamiltonian function governing the dynamics of the planetesimal $m$ is given by
\be
\label{eq9}
F(L,\Lambda,G-L,\lambda,\lambda_B,\varpi) = -\frac{{\cal G} m_A}{2L^2} + n_B \Lambda - R
\ee
where $n_B$ is the mean-motion of the perturbing mass $m_B$ and $R$ the disturbing function. We can express this Hamiltonian in a form adequate for perturbation theory: $F = F_0 + \varepsilon F_1$, where
\bea
\label{eq10}
F_0 &=& -\frac{{\cal G} m_A}{2L^2} + n_B \Lambda \\
F_1 &=& -\frac{a_B}{|{\bf r} - {\bf r_B}|} + \frac{r a_B}{r_B^2} \cos{\phi},    \nonumber
\eea
and $\varepsilon = {\cal G} m_B/a_B$ is a small parameter that serves as a guide of the relative magnitudes between the perturbation term $F_1$ and the unperturbed integrable Hamiltonian $F_0$. 

For the disturbing function, we adopt a Legendre expansion, truncated to fourth order in the ratio $a/a_B$; in other words, we approximate the perturbation by
\be
\label{eq11}
F_1 = \sum_{i=2}^4 \biggl( \frac{a}{a_B} \biggr)^i \biggl( \frac{r}{a} \biggr)^i 
      \biggl( \frac{r_B}{a_B} \biggr)^{-(i+1)} P_i (\cos{\phi}),
\ee
where $P_i(\cos{\phi})$ is the Legendre polynomial of degree $i$. Switching from a power series in $\cos{\phi}$ to a harmonic decomposition in $\phi$ and transforming them to orbital elements, we can obtain a truncated expansion of the disturbing function leading to
\be
\label{eq12}
F_1 = \sum_{i,j,s=0}^\infty \sum_{k,l=- \infty}^\infty D_{i,j,k,l} e^i e_B^j
 \cos{(k M + l M_B - s\varpi)}
\ee
where $M$ and $M_B$ are the mean longitudes of both bodies, and $D_{i,j,k,l}$ may be obtained in terms of the Hansen coefficients (see Beaug\'e and Michtchenko 2003). 

Having an explicit expression for $F_1$ in mean variables, we may now apply Hori's method. The idea is to search for a Lie-type canonical transformation $B = \varepsilon B_1 + \varepsilon^2 B_2 + \ldots$ to a new set of variables $(L^*,\Lambda^*,(G-L)^*,\lambda^*,\lambda_B^*,\varpi^*)$ such that the transformed Hamiltonian $F^*$ is independent of $\lambda^*$ and $\lambda_B^*$. Up to second order in the small parameter, the new Hamiltonian function may be written as
\be
\label{eq13}
F^*((G-L)^*,\Delta \varpi^*;L^*,\Lambda^*) = F_0^* + \varepsilon F_1^* + \varepsilon^2 F_2^*
\ee
where $\Delta \varpi^* = \varpi^* - \varpi_B$. The different orders in expression (\ref{eq13}) are given by
\bea
\label{eq14}
F_0^* &=& F_0 (L^*,\Lambda^*) \nonumber \\
F_1^* &=& \langle F_1 \rangle_{\lambda,\lambda_B} \\
F_2^* &=& \frac{1}{2} \langle \{ (F_1 + F_1^*), B_1 \} \rangle_{\lambda,\lambda_B} \nonumber 
\eea
where $\{ \}$ is the Poisson bracket, $\langle  \rangle_{\lambda,\lambda_B}$ denotes the averaging with respect to both mean longitudes (keeping all other variables fixed), and $B_1$ is the first-order generating function of Hori's method. In terms of the adopted expansion for the disturbing function (\ref{eq12}), it is given by
\be
\label{eq15}
B_1 = -\sum_{i,j,s,k,l} \frac{D_{i,j,k,l}}{k n + l n_B} {e^*}^i {e_B^*}^j \sin{(k M^* + l M_B^* - s\varpi^*)},
\ee
where the function must be evaluated in the new variables. 

The construction of the new secular Hamiltonian $F^*((G-L)^*,\Delta \varpi^*;L^*,\Lambda^*)$ is cumbersome, although fairly straightforward when using an algebraic manipulator. Fortunately, it will not be necessary to write an explicit expression here. Let it suffice to say that $F^*$ constitutes a second-order model of the secular system and a single degree of freedom system in variables $((G-L)^*,\Delta \varpi^*)$. Employing the inverse transformation from Delaunay variables to orbital elements, we can also obtain an expression for $F^*(e^*,\Delta \varpi^*;a^*)$ in terms of the mean eccentricity $e^*$ and the proper semimajor axis $a^*$. Since the latter orbital element is constant, it appears in the Hamiltonian as an external parameter.

Finally, after solving the secular system and obtaining both $e^*$ and $\Delta \varpi^*$ as functions of time, we may invoke the inverse Hori transformation to obtain the short-period variations of the original osculating variables. For the eccentricity, this yields
\be
\label{eq16}
e^2(t) \simeq {e^*}^2(t) + \frac{2 \varepsilon}{L^*}\frac{\partial B_1}{\partial \varpi^*}.
\ee
Because $B_1$ explicitly depends on the mean longitudes, the second term models the short-period variations in the eccentricity, while the first term (${e^*}^2(t)$) gives the main secular contributions. Since the eccentricity is a positively defined function, the magnitude of the second term also specifies the minimum mean eccentricity $e^*$ of the secular system for any given proper semimajor axis $a^*$. At the same time, it also gives the averaged semi-amplitude of the short-period variations $ \Delta e$ in the same orbital element.

\begin{figure}
\centerline{\includegraphics*[width=0.99\columnwidth]{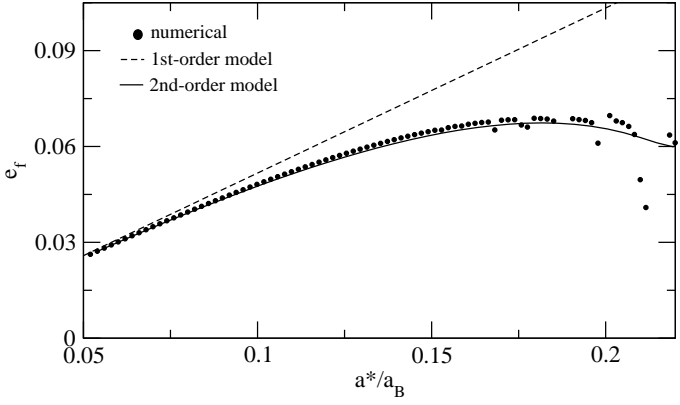}}
\caption{Forced eccentricity as function of the proper semimajor axis, calculated by three different methods: filtered exact numerical simulations (filled black circles), first-order analytical model (dashed lines), and the new second-order secular model (continuous lines).}
\label{fig2}
\end{figure}

Figure \ref{fig2} shows an application of our second-order model to the same generic binary system as was discussed in Figure \ref{fig1}. The plot shows the forced eccentricity, as a function of the ratio $a^*/a_B$, calculated with three different methods. Recall that this model predicts a linear increase of $e_f$ with the semimajor axis. Finally, the value of the forced eccentricity determined with our second-order Hamiltonian $F^*$ is shown as a continuous curve. The agreement with the numerical data is very good, and the saturation in the value of $e_f$ is reproduced quite well. Since we have avoided all small denominators in the generating function $B_1$, the model curve is smooth and shows no indication of the effects of mean-motion resonances.

\subsection{Extending the Th\'ebault et al. (2006) approximation}

As mentioned before, although the second-order model leads to significant improvement in the secular solution, as well as allowing the magnitude of the short-period orbital variations to be modeled, it is much too complex to constitute a workable model. For this reason, we wondered whether the empiric correction term introduced by Th\'ebault et al. (2006) for the secular frequency could be extended to reproduce both the forced eccentricity and the short-period variations. Of course it is not expected to yield the exact same results, but if the errors are not significant, such an empirical second-order approximation could constitute a simple quantitative analytical model.

Following the same approach as Th\'ebault et al. (2006), we use ${e_f}_0$ and $g_0$ to denote the first-order expressions for the forced eccentricity and secular frequency, and reserve $e_f$ and $g$ for the second-order values. The idea then is to write $e_f = {e_f}_0 ( 1 + \varepsilon \delta e_f)$ (and a similar equation for $g$), and attempt to model the correction terms $\delta e_f$ and $\delta g$. After several tests and multivariate linear regressions, we find that the expressions
\bea
\label{eq17}
e_f & \simeq & {e_f}_0 \left[ 1 - 16 \biggl( \frac{m_B}{m_A} \biggr) \left( \frac{a}{a_B} \right)^2 (1-e_B^2)^{-5} \right] \\
g & \simeq & \;\; g_0 \left[ 1 + 32 \biggl( \frac{m_B}{m_A} \biggr) \left( \frac{a}{a_B} \right)^2 (1-e_B^2)^{-5} \right] \nonumber
\eea
agree with the complete second-order model very closely. There are some slight differences in $g$ with respect to the original formula introduced by Th\'ebault et al. (2006) but they are minor and not very significant. Finally, the expressions for ${e_f}_0$ and $g_0$ are those given in (\ref{eq3}) and (\ref{eq4}).

In terms of (\ref{eq17}) the secular Hamiltonian may be approximated well by
\be
\label{eq18}
F^* \simeq n^* {a^*}^2 g \biggl[ \frac{1}{2}({k^*}^2 + {h^*}^2) - e_f k^* \biggr] 
\ee
where $k^* = e^*\cos{(\Delta \varpi^*)}$ and $h^* = e^*\sin{(\Delta \varpi^*)}$ are the new regular secular variables. 

\begin{figure}
\centerline{\includegraphics*[width=0.99\columnwidth]{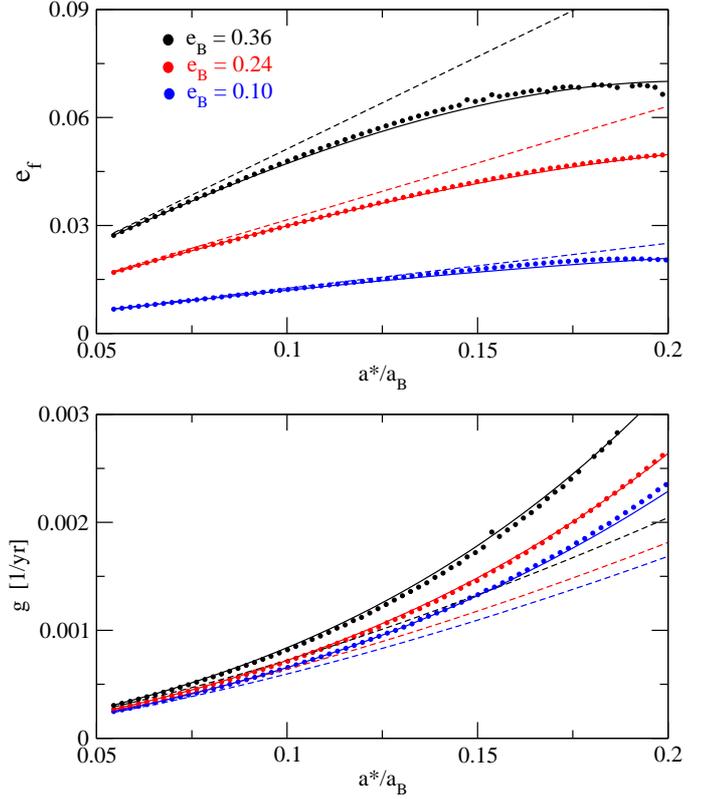}}
\caption{Variation in the forced eccentricity (top) and secular frequency (bottom), in terms of the proper semimajor axis, for three values of the binary eccentricity $e_B$. As before, filled circles present results from filtered exact numerical simulations, dashed lines correspond to the first-order analytical model, while the empirical solutions (\ref{eq17}) are shown in continuous lines.}
\label{fig3}
\end{figure}

Finally, the semi-amplitude of the short-period variations in eccentricity can also be empirically modeled according to the expression
\be
\label{eq19}
\Delta e \simeq 10 \biggl( \frac{m_B}{m_A} \biggr) \left( \frac{a}{a_B} \right)^3 \frac{e_B}{(1-e_B^2)^{6}}. 
\ee

Figure \ref{fig3} once again compares the estimated values of $e_f$ and $g$, this time for three different values of the eccentricity $e_B$ of the binary component. 
 Given the simplicity of these equations, the agreement with the numerical results is surprisingly good.

\section{The $\gamma$-Cephei binary system}

After specifying the basic ingredients of our second-order dynamical model, we attempt to apply it to $\gamma$-Cephei. As mentioned in the introduction, this is probably the best-studied tight binary system with a known planetary body. Since the main secular parameters $g$ and $e_f$ strongly depend on the stellar masses and orbital elements of the secondary star, we begin our discussion by reviewing the accuracy of these parameters.

Throughout this work we refer to the more massive stellar component by $\gamma$-Cephei-A, while $\gamma$-Cephei-B is used to identify the less massive star. The giant planet orbiting $\gamma$-Cephei-A is called $\gamma$-Cephei-b. The masses of each body are denoted by $m_A$, $m_B$, and $m_p$, in that order.

\subsection{History and radial velocity data}

Several years before the discovery of the first planetary body around a main sequence star (Mayor and Queloz 1995), Campbell et al. (1988) suggested the presence of a Jupiter-mass object in a $2.7$ yr orbit around $\gamma$-Cephei-A. The authors, however, remained cautious about claiming a true planetary detection, since the observed periodic variations in radial velocity (RV) were at the very limit of the instrumental resolution. To complicate the problem even further, the variations in RV attributed to the Jovian planet, with a semi-amplitude of only about $25$ m/s, were superimposed on a much larger variation caused by a previously unnoticed stellar companion with a much longer orbital period.

The planetary interpretation was questioned later when changes in the chromospheric activity were observed with similar period (Walker et al. 1992). Thus, it was proposed that the observed changes in RV were spurious and probably only due to changes in the spectral line profiles caused by surface inhomogeneities (spots).

The existence of a binary component (i.e. $\gamma$-Cephei-B) was only reevaluated several years later, when Griffin et al. (2002) combined several historical sources of radial velocities that include epochs from 1896 to 1980. This data set consisted of 88 RV observations, although many of them did not contain proper uncertainties, and a gap of some 50 years was present in the data set. Even so, the authors proposed a secondary stellar mass in the system with an orbital period of $P \sim 66$ yrs.

The presence of a third body, this time a planet around $\gamma$-Cephei-A, was only confirmed by Hatzes et al. (2003) after incorporating new high-precision velocity observations from the McDonald Observatory. They show convincingly that the $2.5$ yr variation in radial velocity was coherent in phase and amplitude throughout the entire $20$ yr interval, as would be expected for Keplerian motion, and that no changes were observed in the spectral-line bisectors.

\begin{figure}
\centerline{\includegraphics*[width=1.0\columnwidth]{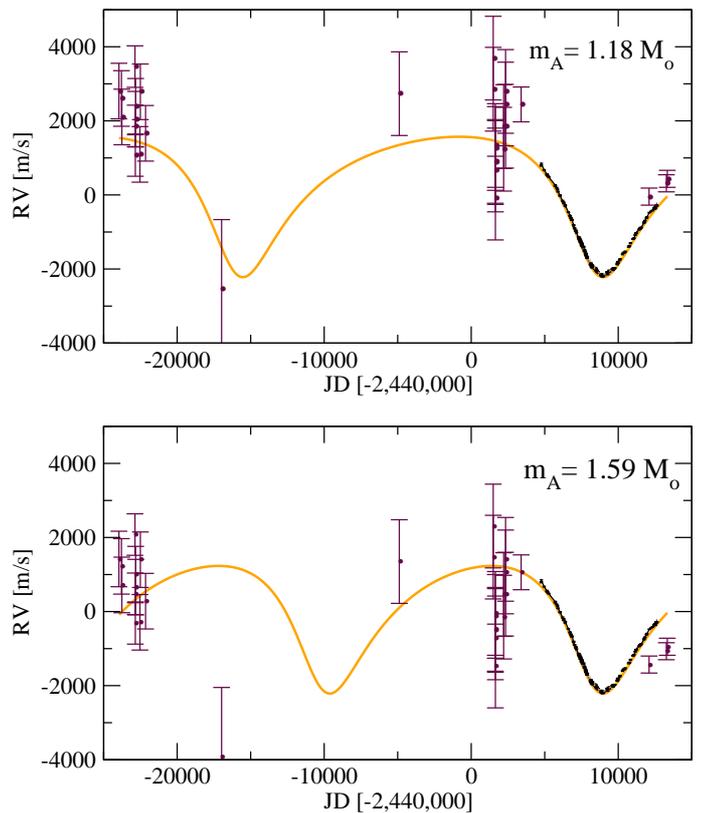}}
\caption{The five sets of RV data used for the orbital fit of $\gamma$-Cephei-B. The four datasets from Hatzes et al. (2003) are shown in black, while the dataset from CfA by Torres (2007) is shown in gray. The error bars correspond to the observational uncertainties given by the authors. Two orbital solutions are shown, one corresponding to a larger orbit for the binary (top) while the bottom panel represents a more compact configuration. Each plot also assumes a different value for $m_A$.}
\label{fig4}
\end{figure}

More recently, Torres (2007) has again analyzed the historical sources of radial velocities using the extensive Harvard-Smithsonian Center for Astrophysics (CfA) database consisting of $\sim 250,000$ spectra. Torres pointed out that some of the historical radial velocities showed large internal discrepancies when compared with other data taken at similar times and were consequently not reliable. The author constructed a reliable data set consisting of 30 RV observations. The complete sets of radial velocities (four sets by Hatzes et al. (2003) and one by Torres (2007)) are shown in Figure \ref{fig4}, where the errors bars indicate the uncertainty on each numerical value. The difference in precision is remarkable, showing how the incorporation of modern techniques in RV measurements lead to the detection of the planetary mass. This increase in precision also allowed the mean anomaly $M$ and longitude of pericenter $\varpi$ of the binary component to be accurately defined.

Independently and without attempting to identify any planetary body, Gontcharov et al. (2000) studied the mass and orbital parameters of the binary system using astrometric observations from several sources. They obtained an orbital period of $\sim 45$ yr and a total mass of $3 M_{\odot}$. Unfortunately, the individual masses were not specified. In his work, Torres (2007) also combined a total of $140$ astrometric measurements obtained between the years 1898 and 1995. Because of the relatively short time span of these observations compared to the binary orbital period, Torres noted that there is a high probability that part of the orbital motion of the binary has been absorbed into the proper motion components reported by Hipparcos. The astrometric information from both these papers is quite different (compare Fig. 6 in Torres 2007 with Fig. 5 in Gontcharov et al. 2000) and the discrepancy in the data prior to 1979 prevent us from using any astrometric data in our analysis of the orbital determination.

\begin{table}
\caption{Published masses and orbital parameters for $\gamma$-Cephei}
\begin{center}
\begin{tabular}{|l|r|r|r|r|r|r|}
 \hline 
    &   $m_A$    &  $m_B$  & $a_{B}$ &    P    & $e_B$ & $\varpi_B$ \\
    & $[M_\odot]$ & $[M_\odot]$ &  [AU]     & [years] &     &  [deg]   \\
\hline 
(1) &  \multicolumn{2}{c}{$m_A+m_B$ = 3} & -- &   $\sim$ 45  &   -   &   - \\
(2) &   1.59    &   0.34     &   18.50    &   56.81   &   0.36     &   158.8 \\
(3) &   1.18    &   0.36     &   19.02    &   66.8    &   0.41     &   160.9 \\
(4) &   1.40    &   0.41     &   20.18    &   67.5    &   0.41     &   161.0 \\
\hline
\end{tabular} 
\end{center}
\footnotesize{(1) Gontcharov et al. 2000, (2) Hatzes et al. 2003, (3) Torres 2007, (4) Nehusauer et al. 2007.}
\label{table1}
\end{table}

Although the high-precision RV measurements from Hatzes et al. (2003) give very good definitions of the mass and orbital parameters of the planetary body, the orbit of binary itself is far from established. In Table \ref{table1} we summarize the results of four different best fits: not only are there noticeable differences in the semimajor axis, but the stellar masses also show large discrepancies.

The brightness of $\gamma$-Cephei has made it an easy target for spectroscopic studies to determine the effective temperature of the star. This parameter, together with the absolute magnitude, are used to determine its mass. However, the effective temperature for the star varies from 4300 to 5100 K (Torres 2007) yielding a wide variety of possible solutions.

\subsection{Possible orbital solutions for $\gamma$-Cephei}

Together with the RV data, Figure \ref{fig4} also shows two different orbital fits for the binary system, each constructed for different $m_A$. The top frame adopts the value given by Torres (2007), which is significantly lower than the one employed by Hatzes et al. (2003), shown here in the bottom graph. We recall that the higher mass is usually used for dynamical studies in this system.

From the raw RV data, we redetermined the best fits for each value of $m_A$. To this end we used the {\it PISA} code (Pikaia genetic algorithm + simulated annealing) that we developed for our studies of resonant exoplanetary systems (e.g. Giuppone et al. 2009). The minimization procedure implies a determination of ten free parameters: five for orbital parameters and five for the RV offsets. We neglected the presence of the planetary body, since it does not introduce any significant effect on the orbital calculation for the binary system. The values of $m_B$, $a_B$, and $e_B$ obtained for each solution are
\bea
\label{eq20}
m_A &=& 1.18 M_\odot \; {\rm :} \hspace*{0.3cm} m_B = 0.31 M_\odot 
   \hspace*{0.2cm} a_B = 18.84 {\rm AU} \hspace*{0.2cm} e_B = 0.41 \\
m_A &=& 1.59 M_\odot \; {\rm :} \hspace*{0.3cm} m_B = 0.37 M_\odot 
   \hspace*{0.2cm} a_B = 20.70 {\rm AU} \hspace*{0.2cm} e_B = 0.41 \nonumber.
\eea
As mentioned previously, the mean anomaly and the longitude of the pericenter are well specified, and show little change in both fits. We assumed an edge-on configuration. This is an important restriction, since any inclination of the orbital plane of the binary would lead to much higher values of $m_B$ (see Hatzes et al. 2003, Haghighipour 2006). Although both fits correspond to significantly different dynamical systems, they yield practically the same residuals: $(\chi_\nu^2)^{1/2}=1.98$.

From the detection of exoplanets from RV data, we have learned (the hard way) that the best fit obtained with a limited data set does not necessarily correspond to the real configuration of the physical system. This is especially true for a small number $N$ of data points, or when they only cover a fraction of the orbital period. In other words, even if we specify a value for $m_A$, the real mass and orbit of $m_B$ does not necessarily have to coincide with the best fit. 

A way to estimate the possible range of solutions compatible with the observational data is to analyze the shape of the residual function $\chi_\nu^2$ for a series of orbital fits around the minimum value ${\chi_\nu^2}_{\rm min}$. If this function shows a steep increase for small changes in the fitted parameters, then we may have a certain confidence that the best fit is probably very close to the real system. However, a shallow minimum could lead to a wide diversity of possible solutions with almost the same value of $\chi_\nu^2$ and, consequently, to different configurations that are statistically indistinct.

\begin{figure}
\centerline{\includegraphics*[width=1.\columnwidth]{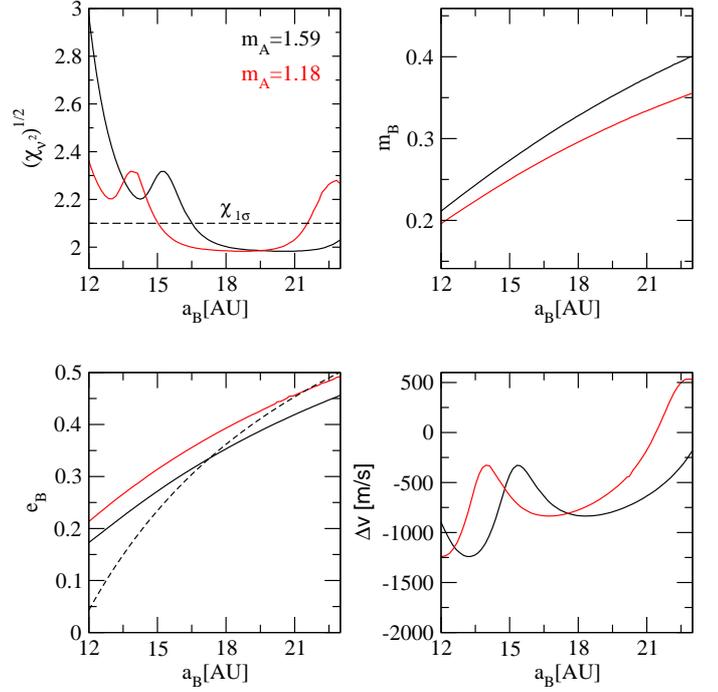}}
\caption{Several multikeplerian orbital fits for $\gamma$-Cephei considering different values of $m_A$ and $a_B$. Black lines show results for $m_A=1.59 M_\odot$, while red curves assume $m_A=1.18 M_\odot$. The dashed horizontal line in the upper-left panel corresponds to the $1 \sigma$ confidence level around the global minimum of $\chi_\nu^2$.}
\label{fig5}
\end{figure}

To test this idea for both choices of $m_A$, we calculated a series of orbital solutions with predefined values of $a_B$ between $12$ and $25$ AU. Except for the pair $(m_A, a_B)$, all the remaining parameters were free and determined with the minimization process. Results are shown in Figure \ref{fig5} for $m_A=1.59 M_\odot$ and for $m_A=1.18 M_\odot$. The value of $(\chi_\nu^2)^{1/2}$ for each fit is shown in the top lefthand plot. Both families of solutions show similar minima, although displaced in semimajor axis. 

To estimate the limits of the $1 \sigma$ confidence level, we employed the same procedure as employed in Giuppone et al. (2009). Given a best-fit algorithm with $\nu$ degrees of freedom, the value of $(\chi_\nu^2)^{1/2}$ associated to the $1\sigma$ confidence level is approximately given by
\be
\label{eq21}
(\chi_\nu^2)^{1/2}_{1\sigma} \simeq (\chi_\nu^2)^{1/2}_{\rm min} 
                           \biggl( 1 + \sqrt{{1 \over 2\nu}} \biggr) .
\ee
where $(\chi_\nu^2)^{1/2}_{\rm min}$ is the minimum value. Since our problem contains $230$ data points and $10$ free parameters, we have $\nu = 220$. For $(\chi_\nu^2)^{1/2}_{\rm min} \simeq 2.0$, equation (\ref{eq21}) then gives $(\chi_\nu^2)^{1/2}_{1\sigma} \simeq  2.1$. 

Although both values only differ in $\sim 5 \%$, the range of possible solutions with $(\chi_\nu^2)^{1/2} \le (\chi_\nu^2)^{1/2}_{1\sigma}$ is surprisingly wide. Assuming the  lower value for $m_A$, the semimajor axis may be as small as $15$ AU or as large as $22$ AU. For $m_B = 1.59 M_\odot$, the range is $a_B \in [17,23]$ AU. 

The top righthand plot of Figure \ref{fig5} presents the values of $m_B$ that give the best fit for each value of $a_B$. As expected, when the components are more separated, the mass of the binary increases to produce same magnitude in radial velocity. The lower lefthand plot gives the binary eccentricity $e_B$. Larger semimajor axes are accompanied by more elliptic orbits. The curves show a rough resemblance to the locus of constant pericentric distance $p_B$, here drawn for one particular orbital solution. Since most of the  RV data points are located near the pericenter of the binary's orbit (see Figure \ref{fig4}), the value of $p_B$ is much better defined than either $a_B$ or $e_B$.

Finally, the lower-right panel shows that the offset from the CfA data varies significantly as a function of the semimajor axis of binary. This behavior is not noted for the other offsets, which remain almost constant. All the best fit solutions showed almost no variation in the longitude of pericenter nor in the time of passage through the periastron.

Table \ref{table2} summarizes the minimum/maximum possible values of $m_B$, $a_B$, and $e_B$ leading to orbital fits with residuals $\chi_\nu^2$ within the $1 \sigma$ confidence level. It is important to stress that statistically all are equally compatible with the observational data. In the next section we attempt to elucidate whether these uncertainties are important to the dynamics of small planetesimals orbiting the primary star and, consequently, whether they could have an effect on the planetary formation process.

\begin{table}
\caption{$1 \sigma$ Confidence limits for $\gamma$-Cephei-B}
\begin{center}
\begin{tabular}{|l|r|r|}
 \hline 
         &    $m_A=1.18 M_{\odot}$   &    $m_A=1.59 M_{\odot}$   \\
\hline 
$m_B  [M_{\odot}]$    &   $[0.25,0.34]$   &  $[0.3,0.4]$  \\
$a_{B} [AU]$               &      $[15,22]$      &  $[17,23]$      \\
$e_{B}$                      &  $[0.32,0.48]$     &  $[0.33,0.46]$  \\
\hline
\end{tabular} 
\end{center}
\label{table2}
\end{table}

\section{Secular dynamics of planetesimals in $\gamma$-Cephei}

\subsection{The forced eccentricity}

Planetary accretion requires low relative velocities which, in turn, implies similar orbits between colliding bodies. This condition may be satisfied if the orbital eccentricities are: (i) very small or (ii) very similar and the orbits are aligned. For small planetesimals where mutual perturbations are not crucial, the orbital eccentricities are determined by a complex interplay between several phenomena, including gas drag, collisional history, and the gravitational effects of the secondary star (e.g. Marzari and Scholl 2000, Th\'ebault et al. 2006, 2008). In the secular approximation, these effects appear through the magnitude of the forced eccentricity $e_f$.

\begin{figure}
\centerline{\includegraphics*[width=0.98\columnwidth]{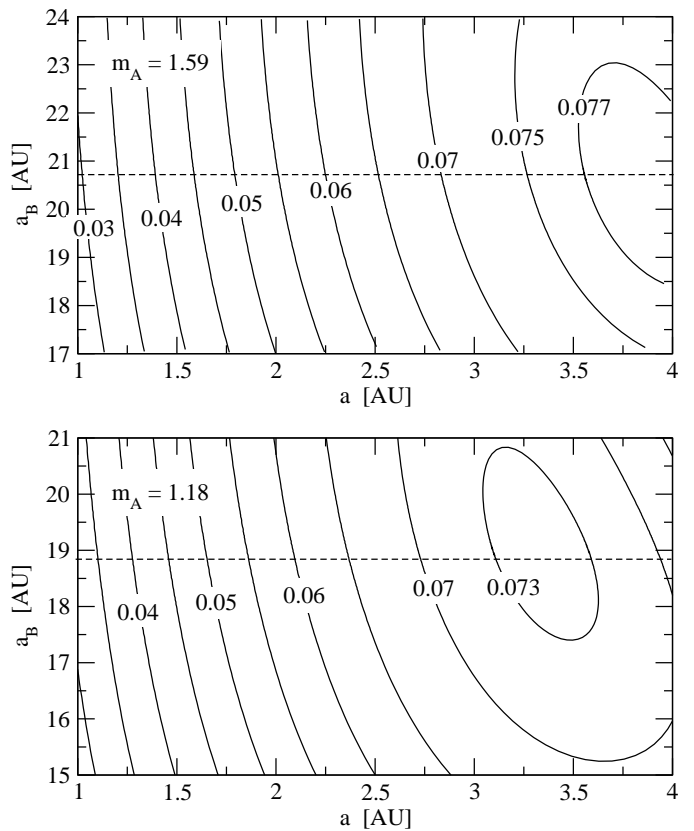}}
\caption{Forced eccentricity $e_f$, as a function of the semimajor axes of the binary $a_B$ and the planetesimal $a$, for all the orbital solutions of $\gamma$-Cephei with residuals $\chi_\nu^2$ within the $1 \sigma$ confidence level of the best fit. Top and bottom frames assume two different values for $m_A$. Each value of $a_B$ implies different values of both $m_B$ and $e_B$, obtained from the families of orbital fits presented in Figure \ref{fig5}. Values of $a_B$ for each best fit are shown with horizontal dashed lines.}
\label{fig6}
\end{figure}

Thus, one way to study planetary accretion under different orbital solutions for $\gamma$-Cephei would be to analyze the sensitivity of $e_f$ to the set $(m_A,m_B,a_B,e_B)$ compatible with the RV data. Results are presented in Figure \ref{fig6} for two values of $m_A$. Each panel shows level curves of constant forced eccentricity, as a function of the semimajor axis $a$ of the planetesimal (abscissa), and for different semimajor axes $a_B$ for the binary pair (ordinate).  

Contrary to our expectations, the range of eccentricities appears insensitive to the binary configuration. In all cases the values of $e_f$ extend from $\sim 0.03$ for the small semimajor axis to $\sim 0.077$ for $a \simeq 4$ AU. Adopting a lower value of $m_A$ seems to cause a slight reduction in the interval, but the change is not very significant. Consequently, and at least from this preliminary analysis, there appears to be no indication that different configurations for $\gamma$-Cephei could cause major changes in the secular dynamics of small bodies and, therefore, on the accretional possibilities of a planetesimal swarm. 

With hindsight, perhaps this result is not at all unexpected. Since all orbital solutions for $\gamma$-Cephei lead to practically the same amplitude in the RV signal, it is understandable that different values for the set $(m_A,m_B,a_B,e_B)$ should also generate similar perturbative effects on other hypothetical bodies in the system; e.g. small planetesimals orbiting $m_A$.

\subsection{Simulations of individual particles with gas drag}

A better test for the effects of different orbital fits on the secular dynamics is to compare the evolution of small planetesimals under the effects of a nonlinear drag force from a circumstellar gas disk centered on $m_A$. We  employ the same expression for the dissipative force as discussed in Beaug\'e et al. (2010). For the gaseous disk we assume a linear surface density profile with a total mass of $3$ Jupiter masses, and an outer edge located at $5$ AU. We consider planetesimals with a volumetric density of $\rho = 3$ g/cm$^3$. 

\begin{figure}
\centerline{\includegraphics*[width=0.98\columnwidth]{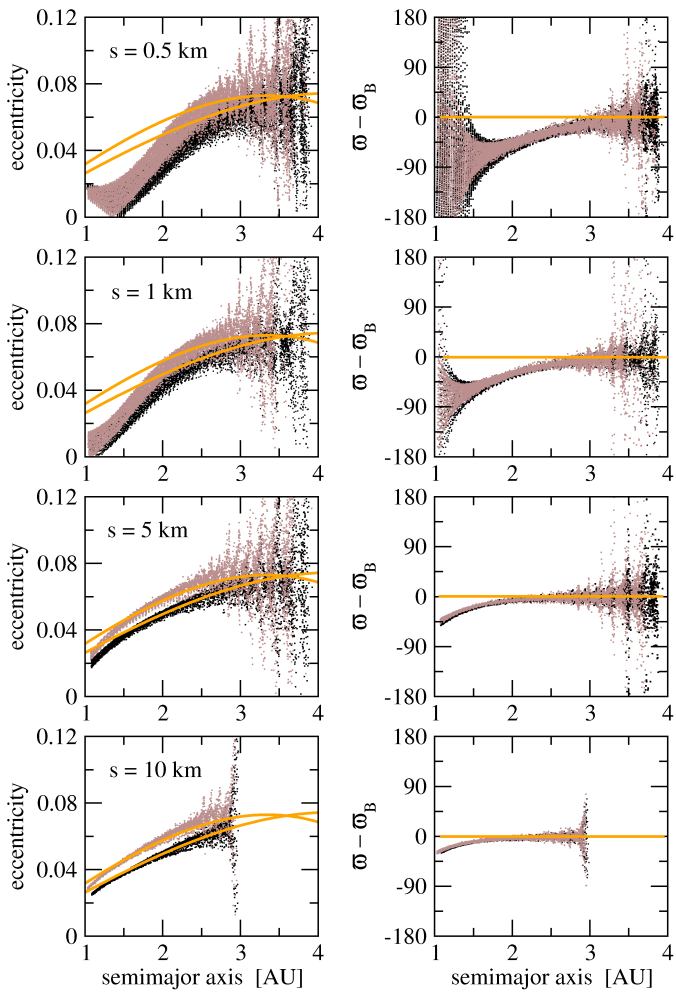}}
\caption{Orbital evolution of four different size planetesimals under the effects of a nonlinear gas drag in the $\gamma$-Cephei system. Black dots correspond to $m_A=1.59 M_{\odot}$ and gray to $m_A=1.18 M_{\odot}$. In both cases the parameters of $m_B$ are those given by the best fits and detailed in equations (\ref{eq20}). The gaseous disk has a constant eccentricity of $e_f=0.05$ and a rigid retrograde precession rate with period $2\pi/|g_g| = 1000$ yr. In the lefthand panels, the orange curve shows the forced eccentricities as a function of the semimajor axis. In the righthand plots, the orange curves marks $\varpi=0$.}
\label{fig7}
\end{figure}

Figure \ref{fig7} shows the result of numerical simulations of four different test planetesimals, with radius between $s=0.5$ km (top) and $s=10$ km (bottom). All were initially placed in circular orbits. The initial semimajor axis was equal to $a=4$ AU for the first three integrations, and $a=3$ AU for the largest planetesimal. We assumed a gas disk with constant eccentricity of $e_g=0.05$ and a rigid-body retrograde precession with a period of $1000$ yrs. 

We considered two different orbital fits for the stellar components. The mass and orbital parameters for the secondary star were taken from  equations (\ref{eq20}) which are the best fit solutions for each value of $m_A$. 

The lefthand panels show the orbital eccentricity as a function of the semimajor axis. The orange curves indicate the forced eccentricity $e_f$, as obtained from our second-order model, for each value of $m_A$. The righthand plots show the evolution of the longitude of pericenter $\varpi$. The origin was shifted so that $\varpi_B=0$. 

We note the existence of three distinct regions in the semimajor axis domain. For $a \lesssim 2$ AU, the planetesimals show short-period oscillations around equilibrium values $e_{\rm eq}$ and $\varpi_{\rm eq}$. These equilibrium values depend on size. Small planetesimals show  $e_{\rm eq} < e_f$ and $\varpi_{\rm eq} < 0$. Larger bodies, however, appear coupled with the conservative secular solution $(k,h) = (e_f,0)$. 

The second region lies roughly between $2$ AU and $3$ AU, and is characterized by very similar equilibrium values of both $e$ and $\varpi$ for all planetesimals with radius $s \gtrsim 0.1$ km. {Moreover, the equilibrium eccentricities are virtually indistinguishable from the forced eccentricity $e_f$.} If planetary accretion is possible with these disk parameters, this region appears to be the most promising since orbital dispersion is kept at a minimum. {The limit between both regions (here at $a \sim 2$ AU) depends on the surface density profile adopted for the disk. In our simulations we used a linear dependence with the radial distance (Beaug\'e et al. 2010), which translates to high densities close to the central star and low values beyond $3$ AU (see Figure 3 of Kley and Nelson 2008).} 

A third region is located beyond $ \sim 3$ AU. Although the secular dynamics also appear similar for all values of $s$, the simulations show large-amplitude oscillations. These are not only caused by short-period terms but also from several high-order mean-motion resonances between the particles and $m_B$. In the plots these commensurabilities can be seen as spikes where the eccentricity is temporarily excited. Without a detailed analysis, it is not possible to establish whether these commensurabilities will inhibit or favor accretion. Although most of our simulations have shown scattering effects and significant orbital misalignment between resonant and non-resonant orbits, we have also found cases of resonant trapping. This appears to be a high-probability outcome for $s \gtrsim 5$ km. 

\begin{figure}
\centerline{\includegraphics*[width=0.98\columnwidth]{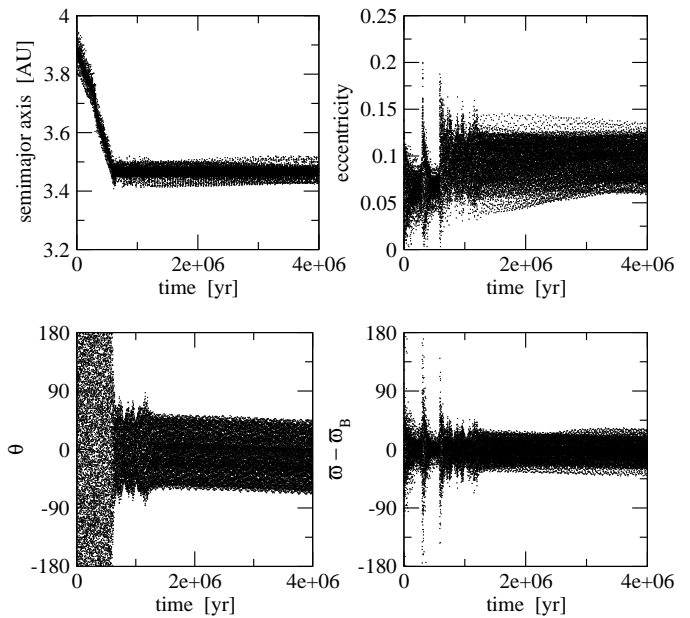}}
\caption{Example of trapping of a $s=10$ km planetesimal in a $10/1$ mean-motion resonance with the secondary star of $\gamma$-Cephei, due to a nonlinear gas drag. Although the orbital migration is divergent, capture still occurs and leads to an apparently stable configuration. The resonant angle is $\theta = 10 \lambda_B - \lambda - 9 \varpi_B$.}
\label{fig8}
\end{figure}

An example is shown in Figure \ref{fig8} for a $10$ km body placed in an initial circular orbit with $a=4$ AU. After an initial decay in the semimajor axis, the body is captured in a $10/1$ mean-motion resonance with the binary component. Both the resonant angle 
\be
\label{eq22}
\theta = 10 \lambda_B - \lambda - 9 \varpi_B
\ee
and the difference in longitudes of pericenter librate around zero, although there seems to be a slow departure towards asymmetric librations at the end of the simulation. The resonant solution seems very stable, at least for timescales between $10^6$ and $10^7$ years. We have found similar outcomes in other commensurabilities, such as the $11/1$ and $12/1$, always for slow dissipative effects (i.e. large bodies). What is curious about this result is that trapping occurs during a {\it divergent} migration; in other words, the non-conservative force increases the separation of the bodies involved. Classical works (e.g. Neishtadt 1975, Henrard 1982, Beaug\'e and Ferraz-Mello 1993, Nelson and Papaloizou 2002) predict that capture is only possible in cases of {\it convergent} migration and, thus, the behavior shown in Figure \ref{fig8} should not occur. 

The only reference we have been able to find describing similar findings is an abstract of a presentation in the 2007 DDA meeting (Hamilton and Zhang 2007). Although no details are available, it appears that divergent trapping is possible in high-order mean-motion resonances and with high-eccentricity perturbers. This may explain why the librating resonant angle includes the longitude of pericenter of the perturber instead of the planetesimal. However, a deeper analysis is necessary before we are able to understand this phenomena and establish its possible importance in planetary formation.

\begin{figure}
\centerline{\includegraphics*[width=0.98\columnwidth]{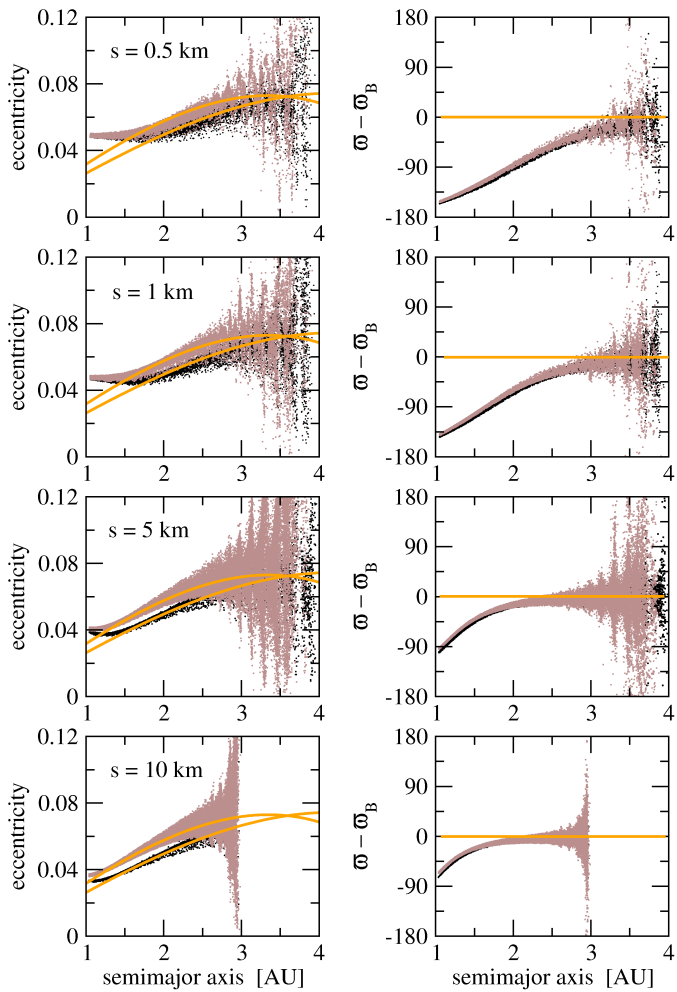}}
\caption{Same as Figure \ref{fig7}, but assuming a static gas disk anti-aligned with the secondary star $m_B$.}
\label{fig9}
\end{figure}

The results shown in Figure \ref{fig7} for both binary configurations show almost identical results. Although the forced eccentricity is slightly higher for $m_A=1.18 M_{\odot}$ for most of the semimajor axis domain, the same three regions exist in both cases, and it is plausible to assume that the collisional evolution of a planetesimal swarm should be similar. Figure \ref{fig9} shows a second series of integrations with the same four planetesimals as before, but this time we considered no precession for the gaseous disk. We also assumed that $\varpi_g - \varpi_B=\pi$, i.e. the disk is anti-aligned with the binary companion. This is consistent with the hydrosimulations of Marzari et al. (2009b) for disks with significant self-gravity. Although the magnitude of the resonant and short-period oscillations appear larger for $m_A=1.18 M_{\odot}$, the averaged secular dynamics is similar in both cases, and the same three regions discussed previously are also present. 

\section{Conclusions}

In this paper we have presented a second-order theory for the secular dynamics of massless particles orbiting a central star and perturbed by a secondary stellar component with high eccentricity. {Only coplanar motion is considered.} This dynamical problem is applicable to the motion of small planetesimals in tight binary systems such as $\gamma$-Cephei. Although the resulting expressions for the forced eccentricity $e_f$ and secular frequency $g$ are complex, we were able to extend the empirical approximation originally introduced by Th\'ebault et al. (2006) and deduce simple analytical formulas for both quantities. 

The forced eccentricity, in particular, shows significant differences to the classical model (e.g. Heppenheimer 1978). While the first-order equations predict a linear dependence with the semimajor axis, numerical simulations show a quadratic functional form, and $e_f$ may actually reach a plateau for high values of $a$. Our model reproduces this behavior with good precision.

We also analyzed the reliability of the best fits presented in the literature for the stellar components of $\gamma$-Cephei. We found that the best solution depends on the adopted mass for the central star, and even for a fixed value of $m_A$ there may be many different configurations compatible with the observations. This is expected, since the radial velocity data covers less than one orbital period of the system. However, we have also found that the dynamics of small planetesimals appears to be only weakly dependent on the particular solution adopted for the binary. A comparative study of the evolution of planetesimals under the effects of gas drag from a circumstellar ($m_A$ centered) gas disk shows similar evolutionary paths. Although our integrations have not covered many initial conditions or disk parameters, we suspect that the accretional process of a planetesimal swarm should be practically equivalent in any case. {Consequently, we believe that the difficulties in explaining planetary formation in tight binary systems cannot be attributed to uncertainties in the orbital fits. The solution must be found elsewhere, and the search is precisely what makes this problem intriguing.}

Finally, we also presented a curious case of resonant trapping in divergent migration. There is practically no reference to this behavior in the literature and, although its importance in planetary formation is difficult to evaluate, we nevertheless believe the phenomena is intrinsically interesting and merits further analysis.

\section*{Acknowledgments}
This work was supported by the Argentinian Research Council -CONICET- and by the C\'ordoba National University -UNC-. A.M.L. is a postdoctoral fellow of SECYT/UNC.

\end{document}